\documentstyle{mn}
\begin{document}

\def\ltsima{$\; \buildrel < \over \sim \;$}
\def\simlt{\lower.5ex\hbox{\ltsima}}
\def\gtsima{$\; \buildrel > \over \sim \;$}
\def\simgt{\lower.5ex\hbox{\gtsima}}
\def\degr{$^{\circ}$}
\input psfig.tex

\title[Reprocessing of radiation by multi-phase gas in Accretion Flows]
{Reprocessing of radiation by multi-phase gas in Low 
Luminosity Accretion Flows} 

\author[A. Celotti and M.J. Rees]
{A. Celotti$^{1,2}$ and M.J. Rees$^2$ \\
$^{1}$S.I.S.S.A., via Beirut 2--4, 34014 Trieste, Italy \\
$^2$Institute of Astronomy, 
University of Cambridge, Madingley Road, Cambridge CB3 0HA}
 
\maketitle
\begin{abstract}
We discuss the role that magnetic fields in low luminosity accretion flows
can play in creating and maintaining a multi--phase medium, and show that
small magnetically--confined clouds or filaments of dense cold gas can
dramatically reprocess the `primary' radiation from tori. In particular,
radio emission would be suppressed by free--free absorption, and an extra
(weak) component would appear at optical wavelengths. 

This is expected to be a common process in various environments in the
central regions of Active Galaxies, such as broad line regions, accretion
disk coronae and jets. 
\end{abstract}


\section{INTRODUCTION}
Observational evidence, from the optical to the $\gamma$--ray band,
suggests the presence of gas with a range of physical properties in the
inner regions of Active Galaxies (AGN). Hot/relativistic material
co--exists with `cold' quasi--thermal gas, on a large range of scales in
these complex environments: tenuous hot plasma forms a corona above or
around relatively cold and dense plasma, in the form of an accretion disk
or re--processing clouds; clumps of photoionized gas, responsible for the
observed narrow and broad emission lines, are embedded in a tenuous hotter
medium. 

The formation, confinement and survival of a multi--phase medium has been
considered with much detail in the case of thermal re--processing by the
material which gives rise to the broad lines. Difficulties associated with
the maintenance of photoionized matter in two phases which are in gas
pressure equilibrium, first led to the suggestion that a more effective
way of sustaining temperature and density gradients may involve magnetic
fields in equipartition with other forms of energy (Rees 1987). 

Magnetic fields are believed to be a common ingredient in the AGN
environment and so can provide effective confinement and insulation of
cold gas embedded in a much hotter ambient medium.  In particular, more
detailed work, focused on smaller scales, has shown that this is plausibly
the case in the magneto--sphere around the central engine (typically on
scales $\leq\ 50 R_{\rm s}$) as well as in relativistic jets, thanks to
the combined action of the magnetic and radiation fields (Celotti, Fabian
\& Rees 1992; Kuncic, Blackman \& Rees 1996;  Kuncic, Celotti \& Rees
1997; Celotti et al. 1998; Celotti \& Rees 1998). 

Recently, much attention has focussed on accretion flows where
electron--ion coupling could be ineffective because the density is low and
the kinetic temperature very high. Very hot ions then provide the dominant
pressure, but radiative cooling is inefficient because the energy cannot
be tapped by the (cooler) electrons (Rees et al. 1982). Accretion flows of
this kind, in which thermal energy is advected inwards rather than being
radiated, may be relevant to the puzzling absence of substantial radiative
dissipation in large elliptical galaxies (Fabian \& Rees 1995).  This has
raised interest in the phenomenon and led to more detailed theoretical
work and the comparison of the expected spectral signatures with data. 
Approximate analytical solutions as well as self--consistent numerical
results for the structure and properties of the flow (re-named ADAF) have
been derived (e.g. Narayan \& Yi 1994, 1995; Chen et al. 1995; Mahadevan
1997; Chen, Abramowicz \& Lasota 1997).  Although the faint ADAF emission
cannot be clearly disentangled from other spectral components over a
significant energy range, the radiative predictions seem to be, so far,
globally consistent with data (e.g.  Lasota et al. 1996;  Reynolds et al.
1996; Di Matteo \& Fabian 1997; Esin, McClintock \& Narayan 1997; Narayan
et al. 1998). However, thanks to very recent observations of a few
elliptical galaxies in the radio to mm frequency range, Herrnstein et al.
(1998) and Di Matteo et al. (1998) have pointed out that (at least)
`standard' ADAF models might be inadequate to account for the (absence of)
emission in large elliptical galaxies. 

Here we suggest the possibility that within the flow some of the gas can
be in the form of cold and dense clumps or filaments, confined by the
magnetic field expected to thread the accreting matter and examine the
effects that this might produce on the observed spectral signatures of
ADAFs. 

\section{The model and its ingredients}

Let us consider the properties of an ADAF, according to the accretion
solutions derived by Narayan \& Yi (1995) and Mahadevan (1997). The
self--similarity postulated by these authors must break down near the
inner boundary, but may nevertheless be an adequate approximation over a
substantial range in $r$. 

The expected field, assumed in quasi--equipartition with the total gas
pressure, is of the order of
\begin{equation}
B=3.7\times 10^2 m_8^{-1/2} \dot m_{-3}^{1/2} r_5^{-5/4} \qquad {\rm G}
\end{equation}
where $m_8=M/10^8 M_{\odot}$ is the mass of the black hole, $\dot m_{-3}
=\dot M/ 10^{-3} \dot M_{\rm E}$ the accretion rate in Eddington units and
$r_5=R/5 R_{\rm S}$ the distance in units of Schwarzschild radii. 
Hereafter, the viscosity parameter and the gas to total pressure have been
taken as $\alpha\simeq 0.3$ and $\beta \simeq 0.5$, respectively, though
it is straightforward to scale our results to other values (the constants
$c_1$ and $c_3$ of the order of unity as defined in Narayan \& Yi 1995,
are taken $c_1=0.5$ and $c_3=0.3$). 

\subsection{Dense and cold gas}

The magnetic field can trap some plasma and, as this is likely to have
cooled to the Compton temperature, a density contrast as high as $\sim
T_{\rm vir}/T_{\rm Compt}$ with respect to the external/hot phase medium
can be attained. Two body processes start to become very efficient in
cooling the dense trapped gas, which can thus cool still further,
typically reaching temperatures of the order of $T \sim 10^4 T_4$ K (see
Section 3.2). 

The gas pressure is limited by the balance with the total magnetic
field stresses and therefore a maximum density for the cold phase,
achieved for a diamagnetic clump of gas, can be as high as 
\begin{equation}
n\simeq n_{\rm hot} T_{\rm vir}/T \sim 2\times 10^{16} 
m_8^{-1} \dot m_{-3} r_5^{-3/2} T_4^{-1}\qquad \rm cm^{-3}, 
\end{equation} 
where $n_{\rm hot}$ represents the density in the hot phase.

\subsection{Geometrical structure}

The spatial distribution of the gas is clearly determined by the field
structure and the net external force acting on it. This would limit the
scale--height of the gas to a typical value such that the tension of the
distorted field (in the perpendicular direction) balances the net effect
of gravity, radiation pressure and acceleration.  Typically, in presence
of loops with a toroidal field component, the action of gravity leads to
dimensions $h_{\rm g}\sim k T/ m_{\rm p}g$.  However, other forces can
dominate in these environments, such as the gas inertia to dynamical
acceleration and the radiation force. Despite the significant opacity
(mostly free--free) of the extremely dense and cold gas, because the low
ADAF luminosity we do not expect radiative acceleration to play a major
role in this environment (see Section 3.2), in contrast with other
situations in the central regions of AGN. 

It should be stressed that, from the point of view of the large--scale
dynamics, these structures of cold and dense gas can be treated as a
uniform loading to the global flow, as long as the individual
clouds/filaments are small enough that, dragged by the field, they are
constrained to follow the same dynamics as the hot material.  This
requires them to be smaller than the scale--height $h_{\rm eff}$ imposed
by the effective local acceleration $a$, which may be dominated by the
effect of magnetic stresses or internal random motions (see also Section
3.1). As these are difficult to quantify, we have a fiducial upper limit
to the typical scale-height of the order of $h_{\rm g}$ \begin{equation}
h_{\rm eff} \simeq 10^6 {g\over a} m_8^{-1} T_4 r_{5}^2 \qquad \rm cm. 
\end{equation}

\section{The survival of the cold phase}

Let us consider the survival of clumps of cold matter thus confined and
advected with the flow. 

\subsection{Diffusive processes in clouds and filaments}

At the high densities inferred, radiative (see Section 3.2)  and
collisional timescales are much shorter that the dynamical one, estimated
as the free fall time, $t_{\rm ff} \simeq 10^4 m_8^{-1/2} r_5^{3/2}$ s,
allowing the gas to reach thermal and radiative equilibrium. Also,
pressure equilibrium with the surrounding medium can be easily maintained
in the global dynamical timescale, for typical dimension of the cold phase
structures smaller than $\simlt 10^{10} m_8^{-1/2} T_4^{1/2} r_5^{3/2} $
cm. 

Although $h_{\rm eff}$ is much smaller that the extension of the hot gas,
it greatly exceeds, for the inferred field strength, the characteristic
gyro--radii of particles in both phases. 

Resistivity is low enough to maintain the coupling between field and gas
on radiative and dynamical timescales.  However, collisions of the hot
particles with the dense clumped gas, can lead to effective thermal
conduction.  As a detailed radiative computation has shown (see Section
3.2), the cold gas is only marginally ionized and, under this
circumstance, the dominant diffusion is due -- at least in the inner ADAF
regions -- to interactions with neutral atoms, with a typical mean free
path $\lambda \simeq 0.1 n_{16}^{-1}$ cm. In a timescale $t$ the cloud
depth affected by thermal conduction, according to the diffusion
approximation, is of the order of $d \sim 4 \times 10^4 n_{16}^{-1/2}
t^{1/2}$ cm. Furthermore, the collision frequency is so high with respect
to the gyration frequency (typically $\nu_{\rm coll} \sim 10^{2} \nu_{\rm
gyr}$) that even in the direction perpendicular to the field line,
conduction is effective (the collisional frequency is proportionately
lower on larger scales). Even when clouds are thinner than the diffusion
length, radiative cooling may allow them to survive. 

In a disc corona or a relativistic jet the field can be either anchored in
the disk (corona) or have an ordered component (jet).  A possibly severe
threat to filament survival -- as already mentioned -- is posed in an
environment where random motions of field loops and large--amplitude
Alfven waves would cause violent local accelerations, which
correspondingly reduce the thickness of filaments that could be carried
with the flow. The acceleration would be largest on the smallest scales.
Let us consider, simply as an illustration, a Kolmogorov spectrum where
the velocities associated with a scale $l$ are proportional to $l^{1/3}$:
the associated accelerations would then depend on scale as $l^{-1/3}$.
This means that the filaments of a given scale can less easily follow the
smaller--scale eddies. On the other hand, stretching of field lines would
decrease the scale of the filaments. 

We draw attention to this issue, emphasizing the uncertainty in the fate
of dense filaments that are advected inwards with the flow towards regions
of progressively stronger field nearer the hole.  One can envisage an
element of gas, trapped within a magnetic tube, starting off in the broad
line region and moving into the denser and more extreme conditions
envisaged in the ADAF.  But can the cool--phase gas survive, or will it be
shredded and homogenized?  It is unclear whether the disordered motions
associated with the magnetic viscosity would shred any filaments into
structures so thin that they cannot survive the effects of conductivity
and diffusion. 

This is an important open question for matter which is coupled to the
field lines.  However, material decoupled from the field (e.g. ejecta from
stars, behaving like diamagnetic clouds) could survive more easily since
it could be squeezed by the surrounding field but would not necessarily
follow the violent random motion associated with the
`$\alpha$--viscosity'. Therefore, irrespective of the answer to the
question posed above, small clouds or filaments of magnetically--confined
cool gas could still have important consequences for the ADAF spectrum, as
shown in the next Section. 

\subsection{Radiative signatures}

Because of the tiny transverse dimensions, the cold matter occupies a
small fraction of the total volume, even though a significant amount of
the total mass inflow might reside in this component if its scale--height
is comparable to $h_{\rm g}$ ($\simeq (T /T_{\rm vir}) R$).  Nevertheless,
because of the extreme high density, this matter can be very effective in
absorbing (mainly through free--free) and re--processing radiation
impinging on it. 

Furthermore, the 3D structure of the field can allow the matter to
accumulate on sheet--like surfaces or filaments in the direction
perpendicular to the net force, distributing it with high covering
factors. Even though the volume filling factor is small, the surface
covering factor can easily be of order unity. 

Because of the comparatively low radiation field in ADAFs, the gas might
cool to temperatures of the order of $10^4$ K at $\sim 5 R_{\rm S}$,
allowing a density contrast between the two phases as high as $10^8$. The
low temperature and high density cause the trapped gas to be an efficient
absorber, even more so as a significant fraction of the ADAF radiation is
emitted as thermal cyclo--synchrotron at low frequencies. 

Therefore, if dense filaments existed within an ADAF, with (at least at
some radii) a substantial covering factor, they could modify the standard
ADAF spectrum. In particular, the radio and mm bump could be absorbed by
free--free (re--appearing in the optical or UV bands). [Furthermore, for
sufficiently ionized gas, the cold phase plasma frequency also exceeds
that of the cyclo-synchrotron peak, raising the possibility of collective
absorption or reflection effects.] The actual result on the radio spectrum
would depend on whether the clouds were just in the inner parts (where
they would absorb only the higher--frequency sub--mm radiation) or mainly
in the outer part (in which case they could absorb/reflect everything),
where the gas would be largely molecular and grain could survive.  We
stress here that this is an interesting possible complication in
confronting ADAF models with observations. 

Under these circumstances, the absorbed radiation can be energetically
important and the spectral effect relevant.  We computed the effective
temperature and conditions reached by the cold gas in thermal and
radiative equilibrium, as well as the reprocessed spectrum from a detailed
radiative transfer calculation, by using the code CLOUDY (it should be
however noted that the code results might be not totally reliable in these
extreme conditions). The local incident spectrum has been assumed to be
that of an ADAF at the considered distance (e.g. Mahadevan 1997),
including the synchrotron, Compton and bremsstrahlung components. 

Even in the inner part of the ADAF ($\sim 5 R_{\rm S}$) the equilibrium
temperature, as already mentioned, only reaches $\sim 10^4$ K and the gas
is largely in neutral form.

In Fig.~1 we show the result of the computation when a covering factor of
$\sim 1$ is assumed. Substantially all of the cloud heating is due to
free--free absorption of the cyclo--synchrotron component, which is
balanced by collisional line and free--bound cooling, the longest
radiative timescale being $\sim 50$ s. The spectral features in the
re-processed spectrum are expected to be broadened and shifted by the high
velocity motion of the gas.  This includes any (weak) iron emission line,
which therefore would be probably undetectable. The column density in
neutral hydrogen can be less than $10^{22}$ cm$^{-2}$, thus affecting only
the soft X--ray emission from the ADAF. 

Although the ionization state and the spectral results clearly depends on
the detailed incident spectrum (e.g. the amount of Comptonized synchrotron
photons), the result shown here is representative of the global effect of
the reprocessing by the cold material. 

\begin{figure} 
\psfig{figure=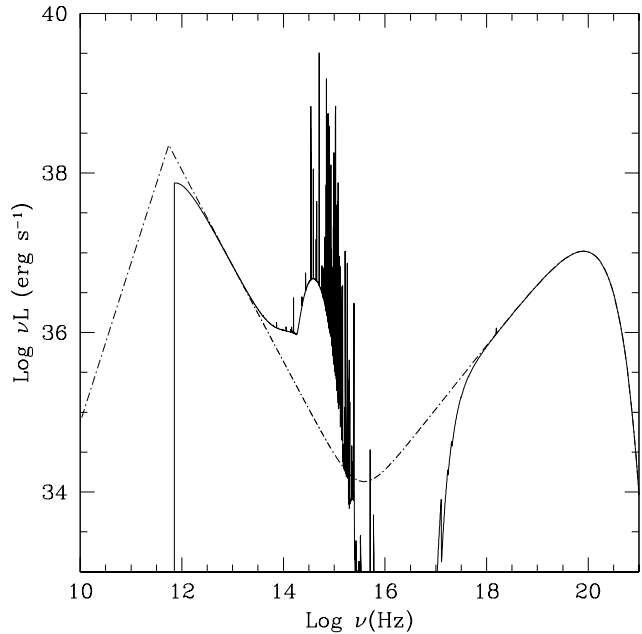,width=0.5\textwidth}
\caption{Reprocessed spectra emitted by cold gas structures 
located in an ADAF environment at about 5$R_{\rm S}$ and with a  
covering factor $\sim 1$. The dashed and continuous lines indicate
the incident and transmitted spectra, respectively.}
\end{figure} 

\section{Discussion and conclusions} 

We have stressed the key role that magnetic fields in quasi--equipartition
in the accretion flow of AGN can have in creating and maintaining a
multi--phase medium, by confining cold, dense gas with small filling
factor.  This is expected to be a common process in the central regions of
Active Galaxies (broad line region, accretion disk coronae, jets,...). 

As mentioned in the introduction, high energy observations of AGN strongly
suggest the presence of gas in different physical conditions in the very
center of AGN. In particular, spectral evidence has accumulated for a
component of optically thick, cold gas, as well as a tenuous hot medium,
to account for the ultraviolet to $\gamma$--ray spectrum of radio--quiet
AGN and galactic black hole candidates.  Interestingly, Krolik (1998) has
recently proposed that a consistent and stable solution for accretion disc
might involve the formation of clumps of dense gas -albeit warmer and
optically thicker than those considered here- embedded in a hot medium and
threaded by magnetic field lines. 

We have shown that thermal gas trapped by the field in an ADAF, can
survive in the hot environment long enough to significantly reprocess the
`primary' radiation and produce observable spectral signatures. Its
predominant effect would be to largely suppress the cyclo--synchrotron
component produced by the hot gas, in the radio--mm band, which has been
considered, so far, an observable indication of the presence of radiative
inefficient accretion phase in our Galactic Centre and in other
low--luminosity galactic nuclei (and Galactic Black Hole candidates). The
resulting spectrum clearly depends on the radial distribution of the
amount and covering factor of the gas. There are here two countervailing
effects that are hard to quantify. The pressures (and hence the maximum
densities of cool gas) are higher at small radii, so a given amount of gas
would cause more absorption near the centre. On the other hand, if the
cool gas consists of filaments entrained in the flow, and advected inwards
with it (rather than being free of magnetic flux and merely squeezed and
confined by the hot--phase medium), the violent random motions associated
with the magnetic viscosity may shred and disperse it before it reaches
the centre: in this case the dominant effect would be due to gas at
100--1000 $R_{\rm S}$. 

Interestingly, the presence of cold clumps might account for the tight
upper limits recently found on the radio, cm and sub-mm emission of large
elliptical galaxies likely to host a massive black hole accreting ISM in
the radiatively inefficient ADAF regime.  In particular the observational
constraints on the 22 GHz emission from the region within the molecular
maser disk in NGC~4258 (Herrnstein et al.  1998) have been strengthened by
the very recent results of VLA and SCUBA observations of three elliptical
galaxies by Di Matteo et al. (1998).  The upper limits found by these
authors are indeed well below the expected emission from `standard' ADAF
models: cold material in the conditions considered here could be
responsible for the absorption of a significant amount of the (unobserved)
ADAF cyclo--synchrotron radiation. 

The strong lines and edges, but especially the thermal continuum shape in
the optical--UV band, expected from the reprocessing by cold material,
might be soon observable, as demonstrated by the upper limits already
derived from the scattered optical light in NGC~4258 (Wilkes et al. 1995). 

\section*{Acknowledgments}
Gary Ferland is thanked, once again, for the use of CLOUDY. The Royal
Society (MJR) and the Italian MURST (AC) are acknowledged for financial
support.


\begin{thebibliography}{}

\bibitem[]{} Celotti A., Rees M.J., 1998, in Theory of Black Hole
Accretion Discs, M. Abramowicz, G. Bj\"ornsson, J. Pringle eds.,
Cambridge University Press, in press

\bibitem[]{} Celotti A., Fabian A.C., Rees M.J., 1992, MNRAS, 255, 419

\bibitem[]{} Celotti A., Kuncic Z., Rees M.J., Wardle J.F.C., 1998,
MNRAS, 293, 288

\bibitem[]{} Chen X., Abramowicz M.A., Lasota J.-P., 1997, ApJ, 476, 61

\bibitem[]{} Chen X., Abramowicz M.A., Lasota J.-P., Narayan R., Yi I., 
1995, ApJ, 443, L61

\bibitem[]{} Di Matteo T., Fabian A.C., 1997, MNRAS, 286, L50

\bibitem[]{} Di Matteo T., Fabian A.C., Rees M.J., Carilli C.L., 
Ivison R.J., 1998, MNRAS, submitted 

\bibitem[]{} Esin A.A., McClintock J.E., Narayan R., 1997, ApJ, 489, 865

\bibitem[]{} Fabian A.C., Rees M.J., 1995, MNRAS, 277, L55

\bibitem[]{} Herrnstein J.R., Greenhill L.J., Moran J.M., Diamond P.J., 
Inoue M., Nakai N., Miyoshi M., 1998, ApJ, 497, L69

\bibitem[]{} Krolik J.H., 1998, ApJ, 498, L13

\bibitem[]{} Kuncic Z., Blackman E., Rees M.J., 1996, MNRAS, 283, 1322

\bibitem[]{} Kuncic Z., Celotti A., Rees M.J., 1997, MNRAS, 284, 717

\bibitem[]{} Lasota J.-P., Abramowicz M.A., Chen X., Krolik J.,
Narayan R., 1996, ApJ, 462, 142

\bibitem[]{} Mahadevan R., 1997, 477, 585

\bibitem[]{} Narayan R., Yi I., 1994, ApJ, 428, L13

\bibitem[]{} Narayan R., Yi I., 1995, ApJ, 444, 231

\bibitem[]{} Narayan R., Mahadevan R., Grindlay J.E., Popham R.G.,
Gammie C., 1998, ApJ, 492, 554

\bibitem[]{} Rees M.J., 1987, MNRAS, 228, 47

\bibitem[]{} Rees M.J., Phinney E.S., Begelman M.C., Blandford R.D., 1982, 
Nat, 295, 17

\bibitem[]{} Reynolds C.S., Di Matteo T., Fabian A.C., Hwang U.,
Canizares C.R., 1996, MNRAS, 283, L111

\bibitem[]{} Wilkes B.J., Schmidt G., Smith P.S., Mathur S., McLeod K.K., 
1995, ApJ, 455, L13

\end{thebibliography}
\end{document}